\documentclass{ws-p8-50x6-00}

\usepackage{times}
\usepackage{amsmath}
\usepackage{amssymb}
\usepackage{mathptmx}

\newcommand{\um}[1]{\ensuremath{\,\mathrm{#1}}} 

\begin{document}

\title{The Time of Flight System of the AMS-02 Space Experiment}

\author{ L.~Brocco, D.~Casadei, F.~Cindolo, A.~Contin, \\
  G.~Laurenti,  G.~Levi, A.~Montanari, F.~Palmonari, \\
  L.~Patuelli, C.~Sbarra, A.~Zichichi}

\address{INFN, Sezione di Bologna, viale Berti Pichat 6/2, 
        I-40127 Bologna, Italy}

\author{G. Castellini}

\address{CNR-IROE, Via Panciatichi 64, I-50127 Firenze, Italy}

\maketitle

\abstracts{The Time-of-Flight (TOF) system of the AMS detector gives
  the fast trigger to the read out electronics and measures velocity,
  direction and charge of the crossing particles.  The new version of
  the detector (called AMS-02) will be installed on the International
  Space Station on March 2004. The fringing field of the AMS-02
  superconducting magnet is $1.0\div2.5$ kG where the photomultiplers
  (PM) are installed.  In order to be able to operate with this
  residual field, a new type of PM was chosen and the mechanical
  design was constrained by requiring to minimize the angle between
  the magnetic field vector and the PM axis. Due to strong field and
  to the curved light guides, the time resolution will be $150\div180$
  ps, while the new electronics will allow for a better charge
  measurement.}

\section{Introduction}

The \emph{Alpha Magnetic Spectrometer} (AMS)~\cite{amsall} has been
redesigned to increase the maximum detectable rigidity up to $1
\um{TV}$, by using a superconducting magnet which will provide a
maximum field of about $0.8 \um{T}$.  The new Time Of Flight (TOF)
system of the experiment has to operate in a stronger magnetic field
and with many different inclinations between the photomultiplier (PM)
axis and the field direction. In the following a description of the
new TOF system is given and results are reported on various aspects of
the system operation.

The main constraints for the TOF sub-detector are due to the operation
in space: the counters must be housed into mechanically robust and
light tight covers; the whole system has to follow the NASA
specifications for the support structure (resistance to load and
vibrations); all the TOF electronics must be protected against
possible faults due to the highly ionized low-orbit environment and
must have complete control on every channel for fine tuning during 3
years of data taking. The power consumption is limited to 150 W for
the whole TOF system.

\section{The new AMS spectrometer and the new TOF design }

\begin{figure}[t]
\begin{minipage}{0.6\textwidth}
\includegraphics[width=\textwidth]{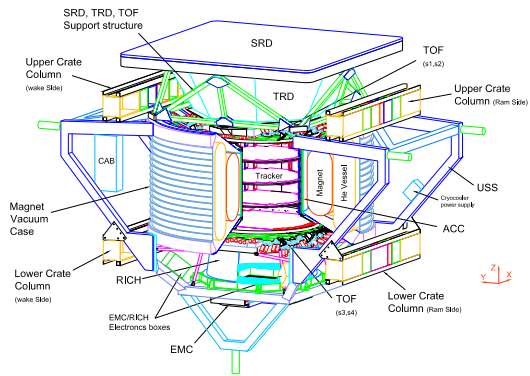}
\end{minipage}%
\hspace{0.05\textwidth}
\begin{minipage}{0.35\textwidth}
\caption{The AMS-02 detector.}\label{expl}
\end{minipage}
\end{figure}

Figure~\ref{expl} shows the new version of the AMS detector (called
AMS-02) that will be installed on the ISS in 2004.  The AMS-02
spectrometer will make use of a superconducting magnet (0.8 T dipolar
field); the tracker will have 8 Si planes, to achieve better rigidity
resolution; there will be a RICH to extend to higher energies the TOF
measurement range, while a TRD and an electromagnetic calorimeter will
improve the capability to distinguish between electrons and protons up
to hundreds GV rigidity.

\begin{figure}[t]
\begin{minipage}{0.55\textwidth}
\includegraphics[width=\textwidth]{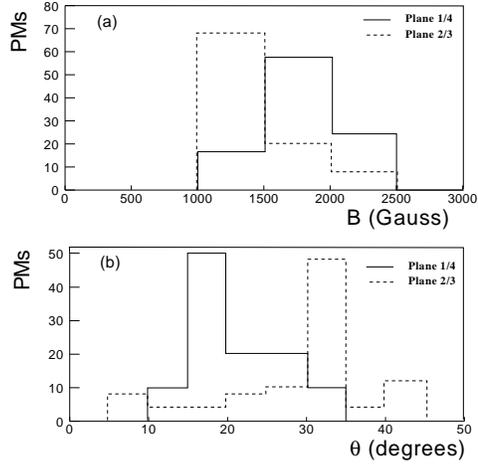}
\end{minipage}%
\hspace{0.05\textwidth}
\begin{minipage}{0.4\textwidth}
\caption{Magnetic field magnitude versus angle $\theta$ 
  between the tube axis and the field direction for planes 1 and 4
  (line) and for planes 2 and 3 (dots).}\label{f:tofangles}
\end{minipage}
\end{figure}

The new TOF is being completely designed and built at the INFN
Laboratories in Bologna. Its main goals are to provide the fast
trigger to AMS readout electronics, and to measure the particle
velocity ($\beta$), direction, position and charge.  The mechanical
constraints of the AMS-02 apparatus do not allow to minimize the PM
orientation with respect to the direction of the magnetic field for
all the TOF counters (see figure~\ref{f:tofangles}). A fine mesh PM
(Hamamatsu R5946), specifically designed to operate in strong magnetic
field, has been chosen for the TOF counters, and thus tested for time
resolution and pulse height response in magnetic field.

\section{Tests in magnetic field of the new TOF phototubes}\label{s:PMs}
 
A red diode has been used to test the response of the PMs, whose light
was guided to the photomultiplier tube by two optical fibres.  The
tube was placed inside the poles of an electro-magnet (maximum field
$4\,$kG) on a movable stand which can be rotated at a maximum angle of
$90^{\circ}$ with respect to the magnetic field.  The charge signal
from the photomultiplier was digitized by an ADC and registered by a
PC-based data acquisition system.

\subsection{PM gain and single photoelectron response}

Three PMs, operating with a gain of $2\times10^6$ at 1700, 2000 and
2200 V respectively, were tested in magnetic field.  The PM responses
have been measured for different values of the magnetic field $B$ and
of the angle between the tube axis and the field direction $\theta$.
The single photoelectron resolution \cite{PMcalib} $\delta$ has been
measured using a very low-level light pulse from the LED, at several
tube orientations and field magnitudes.  It is about 70\% at $B=0$,
but it degrades rapidly with increasing magnetic field at fixed angles
and with increasing angle at fixed magnetic field.  The consequence of
the worsening of $\delta$ is a small degradation in the resolution of
the energy measurement.  No relevant difference is seen between the
three tubes.

\begin{figure}[t]
\begin{minipage}{0.53\textwidth}
\includegraphics[width=\textwidth]{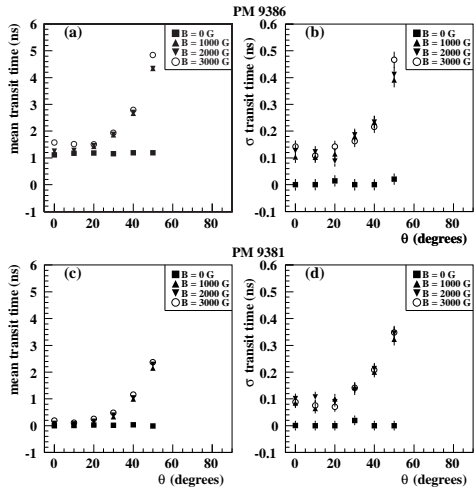}
\end{minipage}%
\hspace{0.05\textwidth}
\begin{minipage}{0.42\textwidth}
\caption{Mean transit time and time resolution as function of $\theta$
  and for different values of B, for PM no.~9386 (a,b) (HV=1700V) and
  PM no.~9381 (c,d) (HV=2200V). For both PM the time at $B=0$ of PM
  9381 is subtracted. The PM operated at highest voltage shows the
  shortest transit time and the best time
  resolution.}\label{f:timediff}
\end{minipage}
\end{figure}

\subsection{Time resolution from data and Simulation of time response}

Figure~\ref{f:timediff} shows the mean transit times (i.~e.~the time
delay with respect to the LED pulse) and the time resolutions for
tubes no.~9381 and 9386, as function of the magnetic field and for
different values of $\theta$ (the times plotted are relative to the
time at $B=0$ of PM 9381).  The most relevant observation is that the
tube operated at highest voltage (no.~9381) shows the shortest transit
time and the best time resolution.  The transit time generally gets
worse with increasing angle for both PMs, but it is more critical for
the PM working at lowest voltage (no.~9386).

The fine mesh PM time response in magnetic field has been simulated by
solving the Lorentz equation of motion for the
photoelectrons,\cite{mioicrc} making use of the Runge-Kutta numerical
approximate solution in finite time intervals.\cite{simfirst} We
followed the energy distribution of the secondary electrons emitted
(SEE) at the dynodes given also by Barbiellini \emph{et al.}
(1995).\cite{simsecond} Finally we got a distribution of time of
arrivals at the anode as a function of $\theta$ and $B$, for different
simulated HV.

\begin{figure}[t]
\begin{minipage}{0.5 \textwidth}
\includegraphics[width=\textwidth]{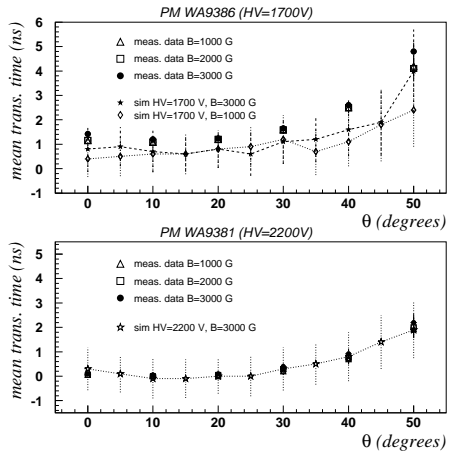}
\end{minipage}%
\hspace{0.05\textwidth}
\begin{minipage}{0.45\textwidth}
\caption{Measured and simulated mean transit time for the PM 9386
  ($-1700$ V) and for the PM 9381 ($-2200$ V) at various $\theta$ and
  fields (the time at $B=0$ of low gain PM is subtracted). The PM
  working at highest voltage shows a shortest transit time also in
  such simulation.}\label{f:simdata}
\end{minipage}
\end{figure}

Figure \ref{f:simdata} shows our simulation of the fine mesh time
response compared to the data (time relative to the time at $B=0$ of
the lowest gain PM). The worsening of the transit time with respect to
the angle $\theta$, both experimental and simulated, is most critical
for the PM working at lowest voltage (no.~9386) and this is also true
for the measured time dispersions (see figure \ref{f:timediff}).

\section{Electronics}\label{s:electronics}

The TOF has 4 planes with 12 counters seen by 2 PMs at each side, for
a total of 192 PMs.  The signals from the two PMs on each side of the
scintillator paddle will be summed to provide one signal from the
anodes and one from the $3^{\rm rd}$ last dynodes.  The anode signals
will be used for time measurement, while the dynode signals will be
used for charge measurement using linear ADCs.  The TOF electronics
will be housed in two pairs of crates, each pair servicing one pair of
planes and containing the PMs power supplies and the read-out
electronics.  

Each electronics crate will feed independently the 24+24 (odd and
even) PMs of two different half TOF planes, in order to guarantee the
necessary redundancy.  In addition the doubly redundant data
acquisition boards will read 24+24 channels (anodes and dynodes) per
crate.

\section{Conclusion}
The AMS-02 TOF system will have a worst time resolution than in
AMS-01, due to the tilted light guides and to the effect of the
magnetic field. In particular, several of the PMs will have an angle
with respect to the magnetic field direction greater than $30^{\circ}$
and thus quite low performances.  The expected resolution in $\beta$
in the worst case, when only two planes are used to compute $\beta$,
would be $\Delta\beta / \beta=3.7\%$ (at $\beta$=1) and positrons
could be identified against protons up to about $1.3\,$GeV.



\begin{thebibliography}{9}


\bibitem{amsall}
The AMS Collaboration,
Phys.~Lett.~\textbf{B 461} (1999) 387-396;
Phys.~Lett.~\textbf{B 472} (2000) 215-226;
Phys.~Lett.~\textbf{B 484} (2000) 10-22;
Phys.~Lett.~\textbf{B 490} (2000) 27-35;
Phys.~Lett.~\textbf{B 494} (2000) 193-202.


\bibitem{TOFpaper}
D. Alvisi \emph{et al.},
Nuclear Instruments and Methods \textbf{A 437} (1999) 212--221.

\bibitem{leo} W.R. Leo, ``Techniques for Nuclear and
  Particle Physics Experiments'', Springer--Verlag (1987).
  
\bibitem{PMcalib} B. Bencheick et al.,
  Nucl. Instrum. and Methods A315(1992) 349.

\bibitem{mioicrc} L.Brocco \emph{et al.},''Behavior of Photomultiplers
  in strong magnetic field for the TOF system of the AMS-02 Space
  Experiment'' ICRC-2001 proceedings.

\bibitem{simfirst} W. Hpress et al.,
  ``numerical Recipes in Fortran 77: The Art of Computing'', Cambridge
  University Press, also avaiable at the url: http://www.nr.com/.
  
\bibitem{simsecond} G.Barbiellini
  \emph{et al.}, ``A simulation study of the behaviour of fine mesh
  photomultipliers in magnetic field '' \textbf{A 362} (1995) 245-252.
  
\end{thebibliography}
\end{document}